\documentclass[conference,10pt,a4paper]{IEEEtran}

\usepackage{makeidx}  
\usepackage{graphicx} \usepackage{epsfig,epsf} \usepackage{subfigure}
\usepackage{enumerate} \usepackage{balance} \usepackage{amsmath}
\usepackage{listings}
\usepackage{color}
\ifCLASSINFOpdf
\else
\fi

\makeatletter

\renewcommand{\@thesubfigure}{(\alph{subfigure})\hskip\subfiglabelskip}
\renewcommand{\@@thesubfigure}{(\alph{subfigure})}
\makeatother

\author{\IEEEauthorblockN{Mali\v{s}a Vu\v{c}ini\'{c}, Bernard Tourancheau, and Andrzej Duda\\}
\IEEEauthorblockN{University of Grenoble, CNRS Grenoble Informatics Laboratory UMR 5217 LIG, Grenoble, France\\
Email: \{Malisa.Vucinic, Bernard.Tourancheau, Andrzej.Duda\}@imag.fr}}

\begin{document}

\title{Performance Comparison of the RPL and LOADng Routing Protocols in a Home
  Automation Scenario}

\maketitle

\begin{abstract} 

  RPL, the routing protocol proposed by IETF for IPv6/6LoWPAN Low Power and
  Lossy Networks has significant complexity. Another protocol called LOADng, a
  lightweight variant of AODV, emerges as an alternative solution. In this
  paper, we compare the performance of the two protocols in a Home Automation
  scenario with heterogenous traffic patterns including a mix of
  multipoint-to-point and point-to-multipoint routes in realistic dense
  non-uniform network topologies. 
  We use Contiki OS and Cooja simulator to evaluate the behavior of the
  ContikiRPL implementation and a basic non-optimized implementation of LOADng. 
  Unlike previous studies, our results show that RPL provides shorter delays,
  less control overhead, and requires less memory than LOADng. Nevertheless,
  enhancing LOADng with more efficient flooding and a better route storage
  algorithm may improve its performance.
\end{abstract} 
%

\section{Introduction}
\label{introduction} 

IETF standardization efforts for sensor networks have enabled their full
interoperability and have made the envisioned ``Internet of Things'' (IoT) a
reality. 6LoWPAN, the IPv6 over Low power Wireless Personal Area Networks
protocol \cite{6lowpan-specs-rfc} is the adaptation layer that allowed tiny
devices to become reachable on the global IP network. 
In parallel, the IETF ROLL working group specified RPL, a Routing Protocol for
Low Power and Lossy Networks (LLNs) \cite{rpl-draft}. 
Many argued that even if RPL met the initial goal of providing the industry
actors with a fully-fledged routing protocol, its significant complexity poses a
threat for implementation interoperability of constrained devices. 
Another routing proposal is LOADng (The Lightweight On-demand Ad hoc Distance-vector routing
protocol - Next Generation) \cite{loadng-draft}, a lightweight variant of AODV
\cite{aodv-rfc}.
The aim of this paper is to analyze the performance characteristics and issues of
the two routing protocols for a specific scenario of Home Automation networks.

Routing in LLNs is one of the key challenges for the IoT emergence. 
The constraints of LLNs have a significant impact on the protocol design. 
Small memory limits the number of stored route entries. Limited energy
supply dictates minimal radio usage and optimized control overhead. The
increasing scale of  IoT networks calls for scalable solutions. 
Finally, Home Automation interactive applications may require low latency
communications. 

Much research work focused on the performance of RPL \cite{rpl-performance-draft, rpl-tiny, rpl-two-cases-studies, rpl-performance} leading
to the common observation that RPL performs well in case of
multipoint-to-point traffic, but induces a large overhead in scenarios where point-to-multipoint traffic is
non-negligible. 
We show how the Contiki RPL implementation behaves in a realistic
Home Automation scenario. 
LOADng is a more recent protocol compared to RPL and less studied. 
Our paper compares both protocols and gives new insights into their performance,
existing issues, and draws possible research directions.

We have used the framework for low power IPv6
routing simulation, experimentation, and evaluation based on Contiki OS and its
Cooja simulator/emulator \cite{contiki-framework}. 
Cooja uses a hardware emulator to run the code, so the simulation results
perfectly reflect the behavior of the underlying protocols.
The only part of the simulation that approximates real-world conditions is the
radio propagation model with the assumption of the Unit Disk Graph. 
We
strictly follow the Home Automation scenario \cite{routingreq-home} by mimicking the logical roles
of nodes and appropriate traffic patterns, as well as by generating realistic
topologies on a virtual home plan.

The paper is organized as follows. Sections \ref{rpl-overview} and \ref{load-overview} give an
overview of RPL and LOADng, respectively. 
Sections \ref{simulation-scenario} and \ref{performance} discuss the simulation
scenario and the performance results. Section \ref{related-work} presents the
related work. 
Finally, we provide some concluding
remarks, observations, and perspectives in Section \ref{conclusion}.

\section{RPL---Routing Protocol for Low Power and Lossy Networks}
\label{rpl-overview}

RPL is a Distance Vector protocol that specifies how to construct a
Destination Oriented Directed Acyclic Graph (DODAG) with a defined objective
function and a set of metrics and constraints. RPL uses a proactive approach:
it finds and maintains routes without any traffic considerations---routes are
created 
even if not used.

RPL  specifies a set of new ICMPv6 control messages to exchange information related to a DODAG:
\begin{itemize}
\item{\textit{DODAG Information Solicitation (DIS)} messages pro-actively
    solicit the DODAG related information from neighboring nodes.}
\item{\textit{DODAG Information Object (DIO)}} defines and maintains upward
  routes.
\item{\textit{DODAG Destination Advertisement Object (DAO)} advertizes prefix
    reachability towards the leaf nodes of a DODAG enabling downward traffic.}
\end{itemize} 
A root starts the DODAG building process by transmitting a DIO. Neighboring
nodes process DIOs and make a decision on joining the DODAG based on the
objective function and/or local policy. A node computes its \textit{Rank} with
respect to the root and starts advertising DIO messages to its neighbors with
the updated information. As the process converges, each node in the network receives
one or more DIO messages and has a preferred parent  towards the
sink. Hence, RPL optimizes the upward routes for multipoint-to-point traffic
that accounts for most of the traffic in LLNs.

To support downward routes, RPL uses DAO control messages that give the prefix
information, the route lifetime, and other information about the distance of the
prefix. RPL RFC \cite{rpl-draft} defines the \emph{storing} and
\emph{non-storing} modes. In the non-storing mode, packets use source-routing
for downward traffic. 
In our study, we focus on the storing mode in which each node keeps track of all
accessible downlink  prefixes. 

The Trickle algorithm \cite{trickle-rfc} governs the emission interval of DIOs. 
The idea is to reduce the control overhead of the protocol by sending DIOs less
frequently when there is no change in the topology. In case of a change in the
network, trickle forces more frequent emissions of DIOs. 
The RPL RFC \cite{rpl-draft} does not specify the mechanisms
for the DAO emission (it is left to implementation). ContikiRPL emits DAOs
with a similar approach to the trickle algorithm based on the DIO
transmission timers.

\section{LOADng}
\label{load-overview}
The LOADng protocol \cite{loadng-draft} uses a reactive approach based on the
idea
that LLNs are idle most of the time so a proactive approach would generate
unnecessary overhead. Thus, LOADng establishes a route towards a given
destination only on demand when there is some data to send. As IETF did a lot of
work on designing AODV, a reactive protocol for MANETs \cite{aodv-rfc},
a logical consequence was to adapt it for LLNs to make it
implementable on memory constrained devices.

When a device has a packet to send towards a given destination, it consults a
routing table and invokes LOADng in case of an invalid entry. 
The protocol floods a \textit{Route-Request (RREQ)} message through
the network to reach all nodes. 
A node receiving a RREQ checks
if it is the message destination. If not, it forwards RREQ to its neighbors. The node also
learns the reverse path towards the originator of the RREQ message and adds it to
the routing table. Eventually, the destination node receives RREQ and
responds by unicasting a \textit{Route-Reply (RREP)} message towards
the request originator. RREP follows the stored reverse
route. At the same time, intermediate nodes learn the forward route towards the
destination. When RREP reaches the request originator, the
bidirectional route is installed at intermediate
nodes. 

One of the main drawbacks of LOADng is the route discovery delay. During the
discovery process, outgoing packets are buffered, which may cause losses in memory
constrained devices. 
Moreover, flooding is highly energy inefficient so nodes may suffer from energy depletion.
Another issue is related to the collisions of control messages due to flooding, which may lead to unnecessary retransmissions.

We have used a LOADng implementation developed in our lab based on the AODV implementation in Contiki.



\section{Simulation Scenario}
\label{simulation-scenario}

Home automation is one of the key applications of the envisioned ``Internet of
Things''. 
Nodes in such networks are typical highly constrained devices. Good examples are
light dimmers, window shades, motion sensors, typical monitoring sensors, remote
control units. When it comes to the network architecture, two approaches are
generally used \cite{routingreq-home}:
\begin{itemize}
\item{\textit{Centralized architecture:} each networked device communicates with
    a central node that controls the network.}
\item{\textit{Distributed architecture:} different devices within the network
    may cooperate to locally control the network.}
\end{itemize}

\begin{figure}[htbp]
\centering
\includegraphics[width=0.7\columnwidth, height=1.95in]{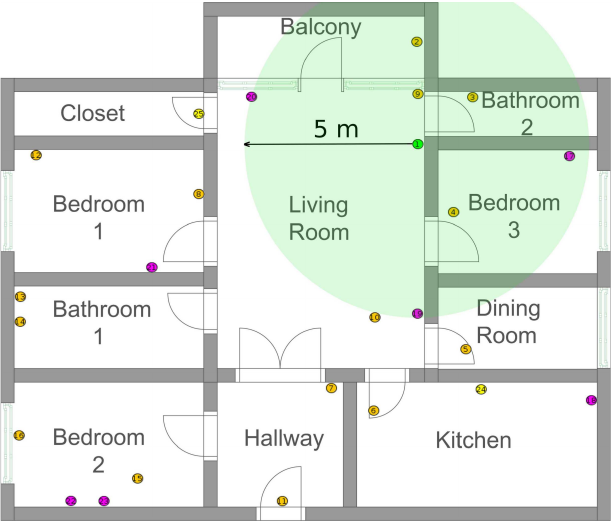}
\caption{An example 25 node topology generated in a virtual house. }
\label{houseplan-25nodes}
\end{figure}

The majority of existing deployments uses the centralized approach, but the
distributed approach is also gaining popularity, mainly because of better
latency. However, in a futuristic scenario, ``smart homes'' are expected to be
a part of a ``smart grid'' and a ``smart city'' with dynamical adjustments of
their energy consumption and production. Thus, the central unit playing the role
of a coordinator among home devices and the external ones, has to exist. For
this reason, we have decided to use the centralized architecture for our
study.

\vspace{-0.2cm}\subsection{Topology}
\label{topology}
A very common simplification in the literature is the assumption of a grid-like
network topology. 
Its main consequence is a near-uniform distribution of the number of node
neighbors (node degree) that does not reflect common ``smart home'' deployment
scenarios. Furthermore, the simplification strongly impacts the performance of
routing protocols. To obtain realistic results, we have designed a virtual
$130\ m^2$ home map and developed a realistic topology generator. 
The home map is the background for generating topologies of different
sizes. Nowadays, networks of around 20 nodes are quite common. 
However, this number is expected to grow and may exceed 250 in the near future
\cite{routingreq-home}. Due to the fixed-size area, topologies with a larger
number of devices in a network become more dense and consequently, the number of
neighbors increases. 

We have generated topologies under the assumption of having 80 \% of devices
uniformly placed on the walls of a room, while the remaining 20 \% are uniformly
distributed around. The number of devices in each room is proportional to its
area. The controlling unit (sink) is placed in the living room. 70 \% of the
devices have the logical role of sensors, while the remaining 30 \% are
assigned the role of actuators. 70 \% of all sensors perform monitoring while the other 30 \%
generate events (switches, motion detectors).

Fig. \ref{houseplan-25nodes} shows an example network topology with 25 nodes in
the virtual house. 
The average number of neighbors for the generated topologies ranged from 4.13
for a 15 node topology, 6.24 for a  25 node network, up to 12.2 for the
largest simulated network composed of 40 nodes.

\vspace{-0.2cm}\subsection{Traffic Pattern}
\label{traffic-pattern}

The traffic pattern of each node in the network depends on its assigned logical
role. We have defined four logical types summarized in Table
\ref{tab:traffic-patern}.
It is important to note that we have chosen the reporting periods and the
general traffic patterns as suggested by IETF
\cite{routingreq-home}. 
Nodes use UDP as the transport layer protocol and the application process
acknowledges each received message. Therefore, traffic in the network is evenly
distributed between point-to-multipoint and multipoint-to-point. 

\begin{table}[htbp]
\centering
    \caption{Traffic pattern of different node types in the network. \label{tab:traffic-patern}
    \vspace{-1.5mm}
    }
    {
    \begin{tabular}{ c p{5.4cm}}
    \hline
    {Node Type} & {Traffic Pattern} \\ \hline \hline
    Monitoring Sensor& Periodic reporting in [8, 12] minute interval \\ \hline
    Event Sensor& Poisson process with a mean of 10 packets per hour for the whole house \\ \hline
    Actuator & Periodic reporting in [8, 12] minute interval and sending acknowledgment frames \\ \hline
    Main Controller Unit & Acknowledgment of received frames; sending a 5 packet burst to
    different actuators as a Poisson process with a mean of 10 bursts per hour and upon reception of a frame from an Event Sensor\\
    \hline
    \end{tabular}
    }
    \end{table}
	
\vspace{-0.2cm}\subsection{Simulation Parameters}
\label{parameters}

To closely reflect real deployments, nodes run Contiki OS with the whole
protocol stack. We have compiled Contiki for the Tmote Sky platform  based on
MSP430 microcontroller with 10 KB of RAM and an IEEE 802.15.4 compliant radio
\cite{tmote-sky}. By using the MSP430 hardware emulator, Cooja thus
takes into account all the hardware constraints of the devices under
study. Table \ref{tab:settings} summarizes the Contiki OS and Cooja
setups.

\begin{table}[htbp]
\centering
\caption{Contiki OS and Cooja parameter setup.\label{tab:settings}
\vspace{-1.5mm}
}
      {
      \begin{tabular}{cc}
        \hline
        {Settings} & {Value}  \\
        \hline
        \hline
        {Wireless channel model} & {UDG Model with Distance Loss}  \\
        \hline
        {Communication range} & {5 m} \\
        \hline
        {Mote type} & {Tmote Sky}  \\
        \hline
        {Transport and network layers} & {UDP + $\mu$IPv6 + 6LoWPAN}  \\
        \hline
        {Max number of queued packets} & {2}  \\
        \hline
        {MAC layer} & {non-slotted CSMA + ContikiMAC}   \\
        \hline
        {Radio interface} & {CC2420 2.4 GHz (IEEE 802.15.4)}  \\
        \hline
        {Simulation time} & {8h}  \\
        \hline
      \end{tabular}      
    } 
\end{table}

Notice a fairly low communication range at the PHY layer. 
Indeed, we have chosen this setup to reflect the assumption that nodes will
operate with a very low power, i.e. with a low power amplifier gain mainly to
reduce interference. Furthermore, radio-propagation obstacles present in a
typical home environment, coupled with 2.4 GHz frequency, limit the
communication range. With our parameters, the whole house is covered
within four hops as suggested before \cite{routingreq-home}. 
We have modified Contiki CSMA to buffer
multicast packets and retransmit them when the channel was found busy at the
initial attempt. Without this modification, the performance of LOADng was highly
degraded due to a large number of dropped RREQ messages. 

Another realistic assumption taken into account in our simulation was the use of
ContikiMAC, a Radio Duty Cycling (RDC) protocol \cite{contikimac} based on
preamble sampling and possibly sleeping receivers. 
Note that ContikiMAC operates on top of the IEEE 802.15.4 non-slotted CSMA and
it may lead to some lost frames, which alleviates the assumption of the Unit
Disk Graph that does not introduce any losses for nodes in the radio range.


\begin{table}[htbp]
\caption{Protocol parameters.\label{tab:params}
}
\subtable[RPL]{
 \begin{tabular}{cp{0.65cm}}
        \hline
         {{DIO} Min Interval (s)} & {4}  \\
        \hline
        {{DIO} Max Interval (s)} & {1048}   \\
        \hline
        {Mode of Operation} & {Storing}  \\
        \hline
      \end{tabular}     
}
\subtable[LOADng]{
 \begin{tabular}{cp{1.2cm}}
      \hline
          {Net Traversal Time (s)} & {10}  \\
        \hline
        {Route Hold Time (s)} &  {600, 1800,}  \\
        {RHT} &  {3600}  \\
        \hline
      \end{tabular}  
      }    
\label{tab:params.}
\end{table}

Table \ref{tab:params} summarizes RPL and LOADng protocol parameters. Notice
that we vary the Route Hold Time (RHT) of LOADng to study the protocol
performance as a function of the route lifetime. 
RPL uses the
Trickle algorithm \cite{trickle-rfc} to emit DIO messages with default
parameters and we have set the DIO Max
Interval used in the steady state to approximately 17
minutes (1048 s). 
Tuning of RHT  and the DIO Max Interval is an engineering
challenge that should take into account the dynamic behavior of a given network,
loss rate, and the probability of a node failure. 
Our choices for the LOADng parameters cover a broad range of cases. 
We can still decrease the overhead of RPL by increasing the DIO Max Interval.

We average the simulation results over 5 simulation runs and show 95
\% confidence intervals. We plot CDF graphs using the cumulative results of 5
simulation runs. 

\section{Performance Evaluation}
\label{performance}

We first focus on a 25 node network (cf. Fig. \ref{houseplan-25nodes}) and
evaluate LOADng and RPL in terms of packet delays, hop counts, and the
routing table size. Furthermore, we study the control plane overhead and the
average routing table size as a function of the number of nodes to show how the
protocols scale for larger
networks.

\vspace{-0.2cm}\subsection{Packet Delay}
\label{delay}

\begin{figure}[htbp]
\subfigure[One hop delays]{
\includegraphics[width=0.46\columnwidth,height=1.3in]{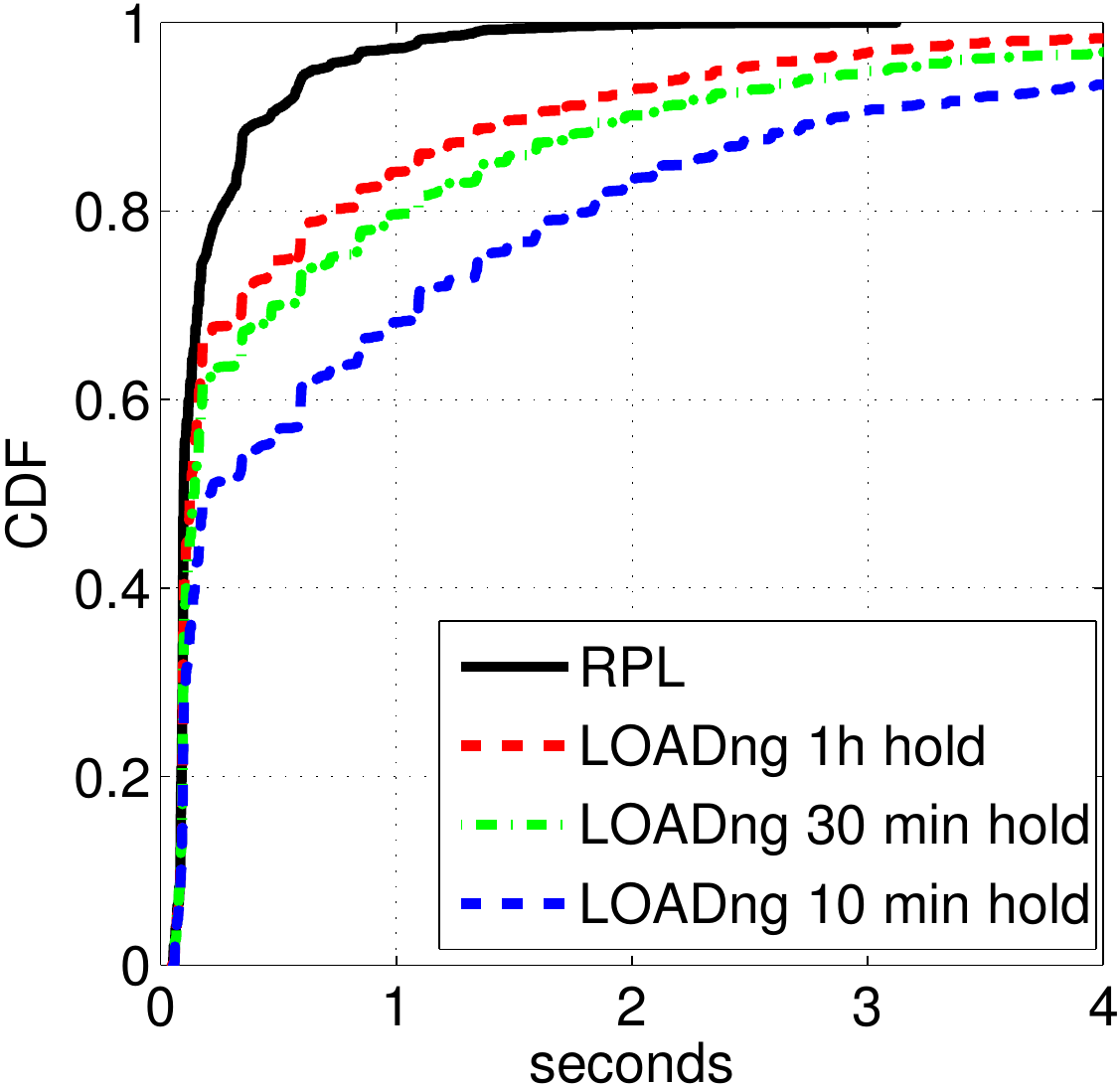}
\label{1hopdelay-cdf}
}
\subfigure[Two hop delays]{
\includegraphics[width=0.46\columnwidth, height=1.3in]{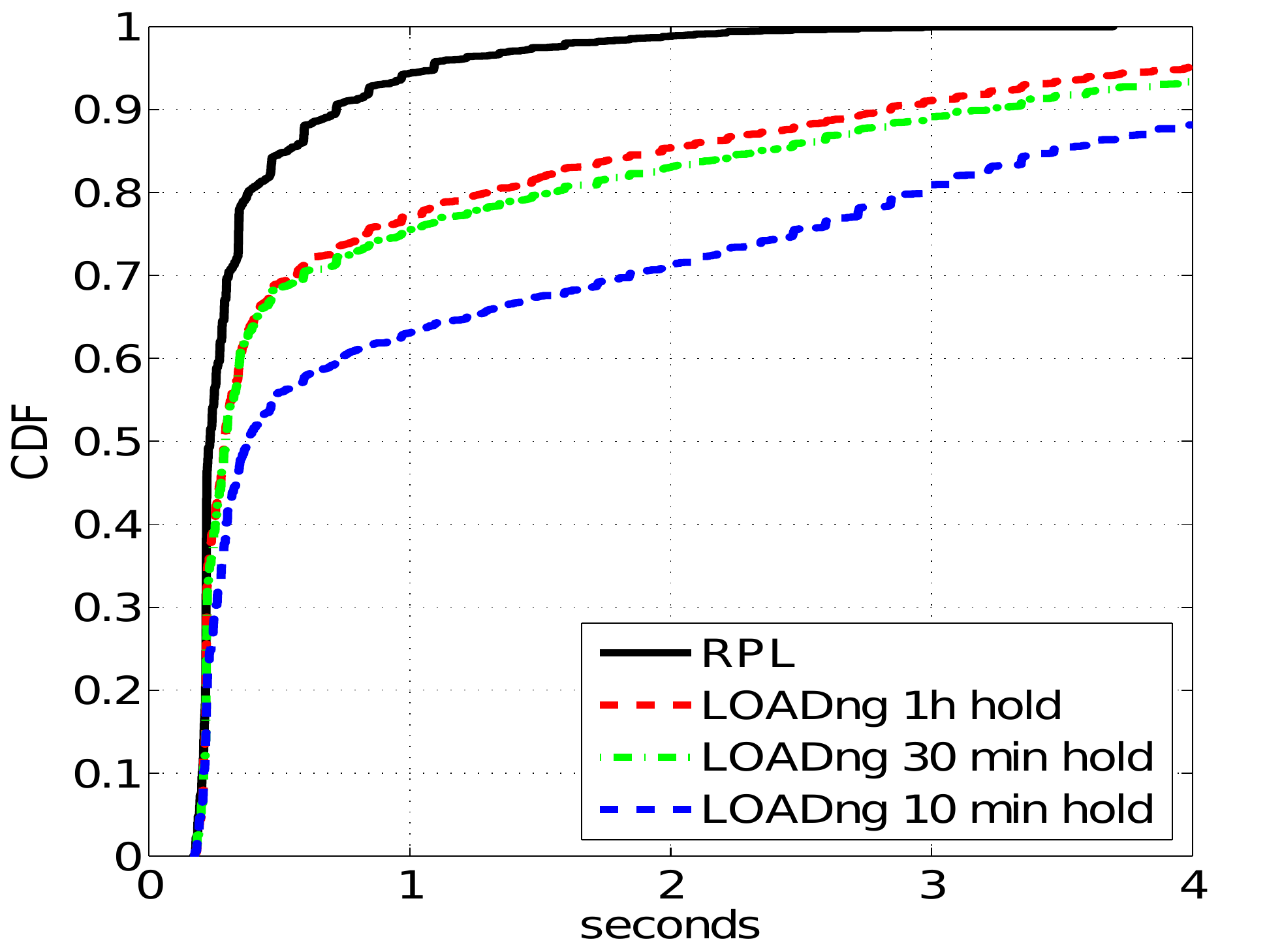}
\label{2hopdelay-cdf}
}
\centering
\subfigure[Three hop delays]{
\includegraphics[width=0.46\columnwidth,height=1.3in]{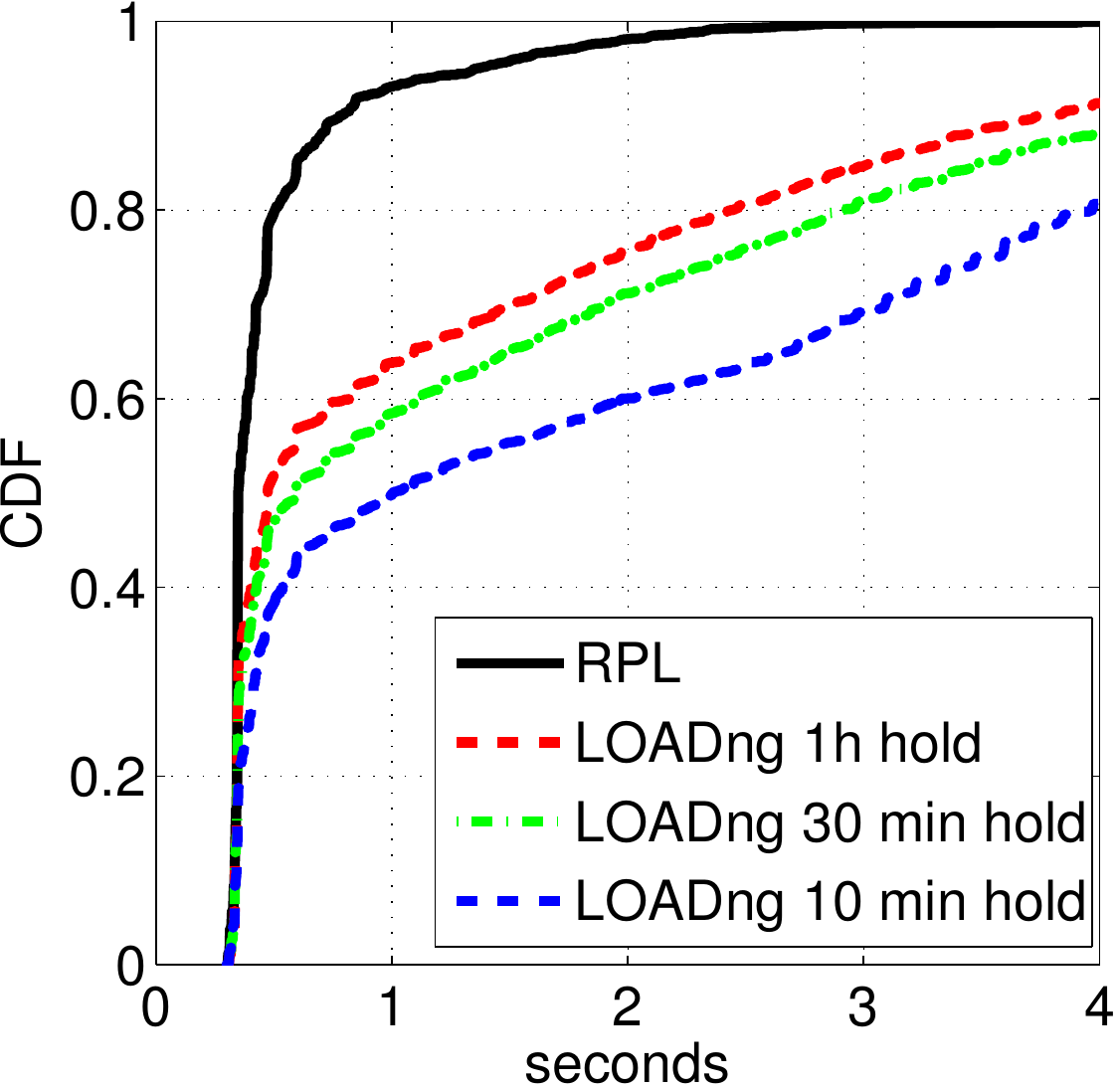}
\label{3hopdelay-cdf}
}
\caption{Packet delay CDFs in a 25 node network.}
\label{fig:hop-delays}
\end{figure}

The latency studied in this section represents the total delay that a packet
experiences from the instant it is passed to the UDP layer until it is
successfully received at the destination. Fig. \ref{fig:hop-delays} presents the
Cumulative Distribution Function (CDF) of packet delays for different hop
counts. 
LOADng performs worse and
packets experience significantly higher delays. Even though some LLNs can be
delay insensitive, the interactive nature of home automation applications
requires a routing protocol with delays of less than 0.5 seconds
\cite{routingreq-home}. Due to the route discovery procedure of LOADng, the
delay CDF converges slowly and 56.8 \% of packets are within 0.5
seconds in case of a 10 minute RHT. Flooding is less frequent when
RHT increases to one hour and the delay  becomes
smaller. However, the delay CDF still converges slowly and approximately 26 \% of
packets experience a delay greater than 0.5 seconds
(cf. Fig. \ref{1hopdelay-cdf}). The pro-active route discovery of RPL results in
shorter delays and 90.9 \% of packets experience a delay less than 0.5 seconds
in case of the one hop distance.

 Packets going through one intermediate node, i.e. two hops away from the sink, (cf. Fig \ref{2hopdelay-cdf})
 experience higher delay, as expected. In case of RPL, 15.4~\% of packets
 experience a delay greater than 0.5 seconds. The maximal delay in this case is
 2.4 seconds. With respect to the 1 hop distance delays, the delay CDF of RPL
 converges slightly slower due to the processing time and the channel contention
 at the intermediate node, because nodes have already a route. 
 In case of LOADng and 1~hour RHT, 31 \% of packets experience a delay
 greater than 0.5 sec.

Moving farther away from the sink does not significantly increase packet delays
of RPL. 
However, we have observed a significant delay increase for LOADng. 
More precisely, 48.2~\% of packets experience a delay greater than 0.5 seconds
(cf. Fig. \ref{3hopdelay-cdf}). In case of a lower, 10 minute RHT,
approximately 62 \% of packets undergo a significant delay. The reason for the
delay increasing with the hop distance is the flooding operation of LOADng as well
as the instillation of forward routes. Basically, a node in the network 
sends a RREQ with the target address of the sink. 
Due to flooding, the sink receives more than one RREQ, each containing different
path metrics. 
It responds with a RREP to the first RREQ and to all other RREQs with lower
metrics. 
Nodes that form the reverse route and forward the RREPs towards its final
destination instill the route towards the sink. 
The sink, however, does not have
a route towards the intermediate nodes and upon the reception of a packet from
them, has to flood the network with another RREQ before sending the
acknowledgment. This results in an increased delay with respect to packets
destined to nodes physically closer to the sink.

\vspace{-0.2cm}\subsection{Hop Distance}
\label{hop-distance} 

We have studied the optimality of the constructed routes through
path hop and packet hop distances. 
The path hop distance represents the average number of hops between a source and
a destination. As the traffic in the 
network is either multipoint-to-point or point-to-multipoint, the sink node is
either a source or a destination of each packet. It is interesting to note that
while RPL has built only bidirectional paths, some paths in LOADng networks are
unidirectional, because of the stochastic nature of the flooding process. 
More precisely, most nodes add a route to the sink during the first
sink-originated RREQ flooding. However, the reverse path is found at a later
time, when the sink needs to address the node. 

\begin{figure}[htbp]
\centering
\subfigure[CDF of path hop distances.]{
\includegraphics[width=0.465\columnwidth,height=1.3in]{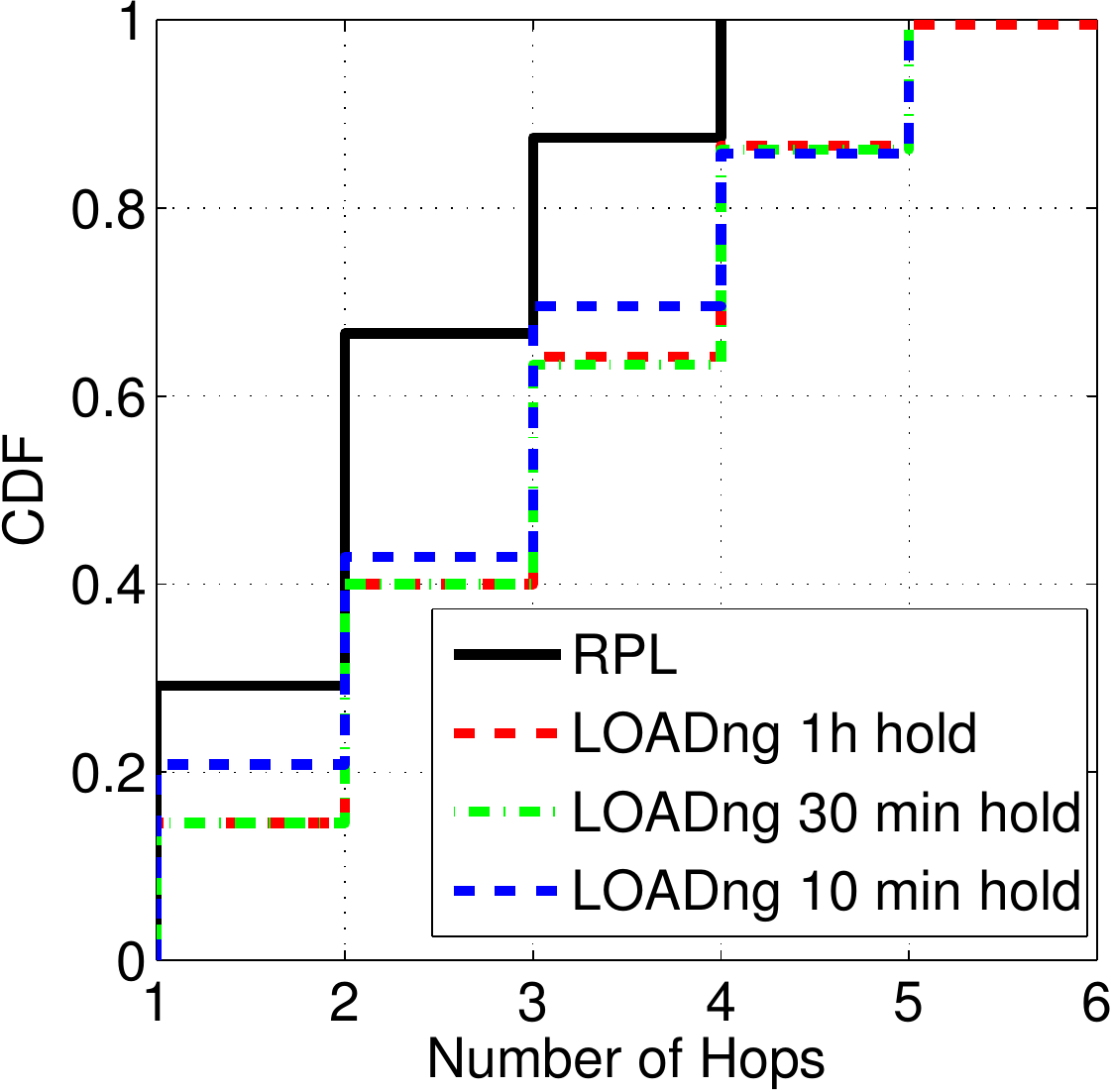}
\label{pathhops-cdf}
}
\subfigure[CDF of packet hop distances.]{
\includegraphics[width=0.465\columnwidth,height=1.3in]{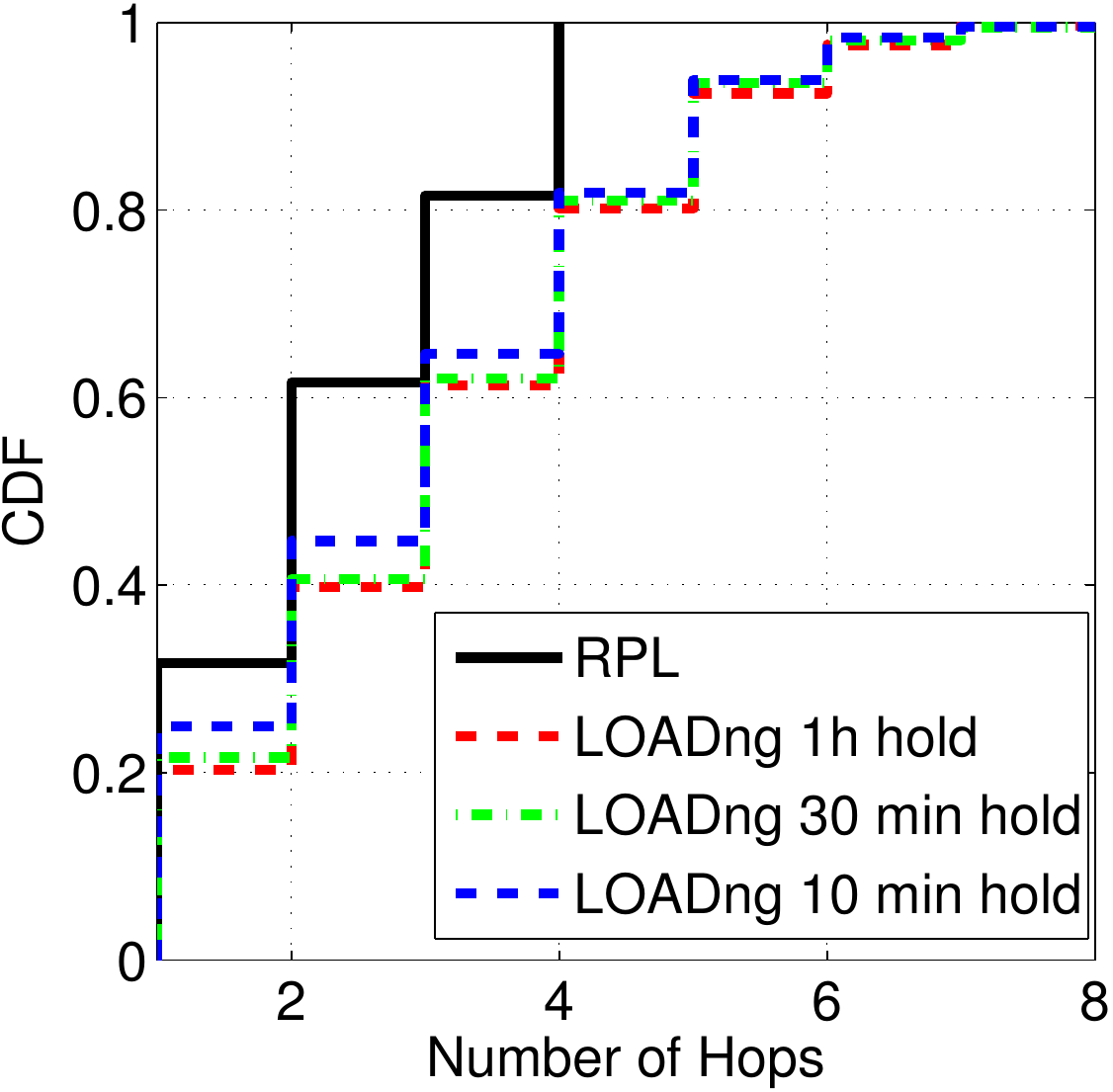}
\label{packethops-cdf}
}
\caption{Distance CDFs in the 25 node network.}
\end{figure}

Fig. \ref{pathhops-cdf} presents the CDF of path hop distances in the network. 
Notice that the RPL routes are shorter than those of LOADng. As mentioned
in Section \ref{topology}, the whole house is covered within four hops. The
figure shows that the average path hop distances of RPL are bounded within four
hops, which is not the case for LOADng. Moreover, 29 \% of RPL routes are within
one hop, while it is around 21 \% for the LOADng, in the case of 10
minute RHT. This gap is even larger for two hop paths: 67 \% of paths in
case of RPL, while only 43 \% in case of LOADng. Note that LOADng provides its
most optimized routes in the case of a shorter RHT. There are two
main reasons for this: i) due to the flooding operation the channel is saturated
and some RREQ messages are dropped, because they reach the maximal number of
attempts of the CSMA protocol, ii) as the nodes are memory constrained, they are
only able to store two packets at a time, so RREQ messages that should be
forwarded are often dropped. As a consequence, more frequent flooding (shorter
RHT) supersedes the lost RREQs and gives more optimized
routes. 

Another issue is the following. If a node has a
packet to send and the route is not ready, the most desired behavior, the
one implemented in our case, is to buffer the packet. Once the RREP arrives, the
node transmits the packet. However, the first arriving RREP is not
necessarily the optimal one in terms of the route length. Therefore, the
packet will follow a non-optimized route. 
We can notice in Fig. \ref{packethops-cdf} that a non-negligible number of
packets go over more than four hops. Upon reception of a subsequent RREP, the
node will 
update its routing table and the following packets will follow the updated
route. 
On the other hand, routes constructed with RPL are optimal: not a single packet
goes over more than 4 hops.

\vspace{-0.2cm}
\subsection{Routing Table Size}
\label{routingtable-size}

\begin{figure}[htbp]
\centering
\includegraphics[width=0.92\columnwidth, height=1.3in]{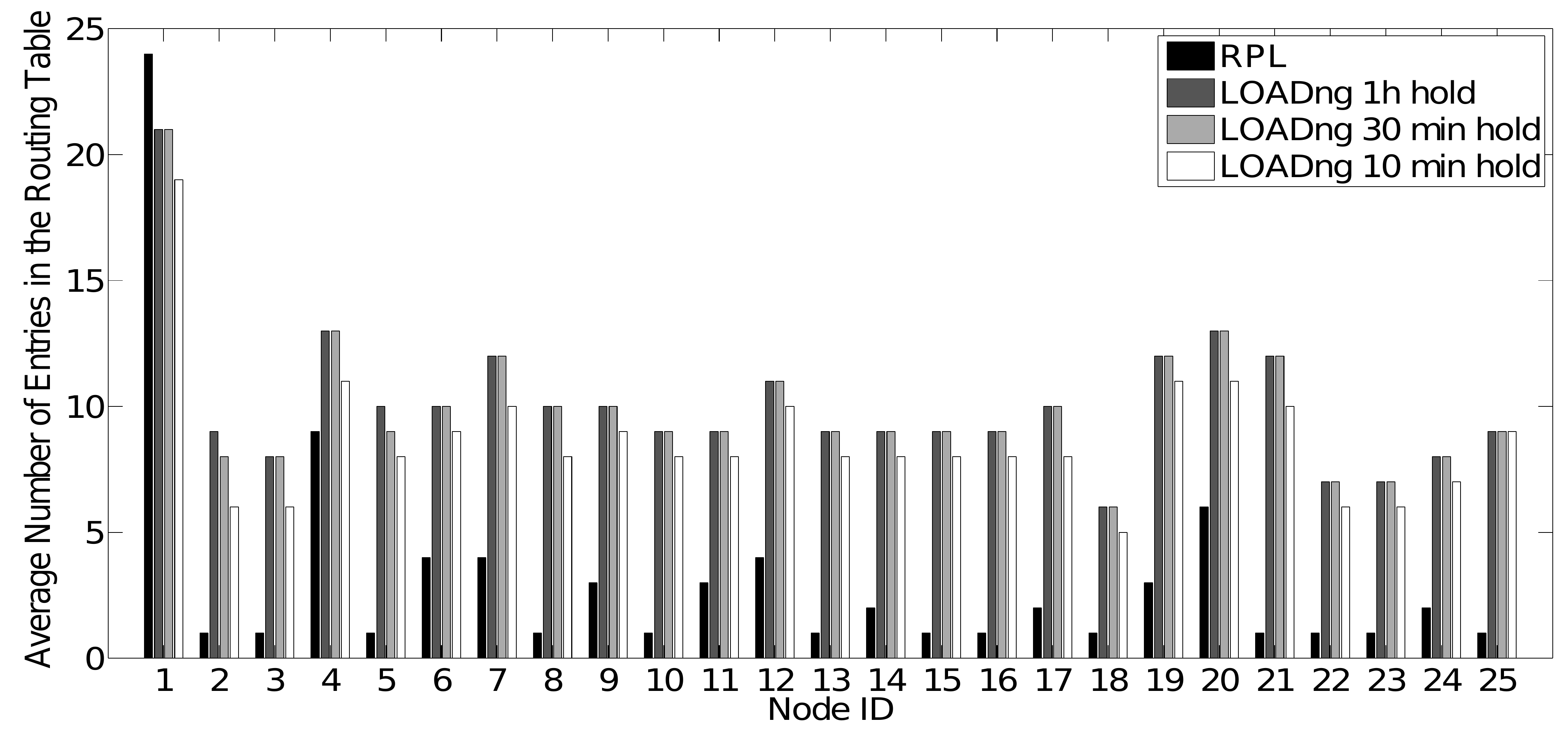}
\caption{Average number of routing table entries during the simulation.}
\label{routingtable}
\end{figure}

Due to the memory constraints of LLN devices, the routing table size is an
important aspect. 
Fig. \ref{routingtable} shows the results for a 25 node network.

LOADng requires much higher number of entries with respect to
RPL. Again, as a consequence of flooding, each node receiving a RREQ or on a
route of a RREP, instills a route towards the sender, which results in a
large number of unnecessary routes. Protocol implementors should thus care
about the priority of RREP routes. This fact could endanger the operation of
the protocol if a node runs out of the available memory. On the other hand, most
nodes in the RPL network just have a default entry towards the preferred
parent. However, depending on their position in the network, some nodes also
have a significant number of entries. Namely, as the RPL operates in the
storing mode, intermediate nodes  selected as preferred parents by
others, have to store downward routes. This is a critical issue, because if
such a node runs out of memory, a loop may be
formed. 

\begin{figure}[htbp]
\centering
\subfigure[Average number of route entries.]{
\includegraphics[width=0.8\columnwidth,height=1.585in]{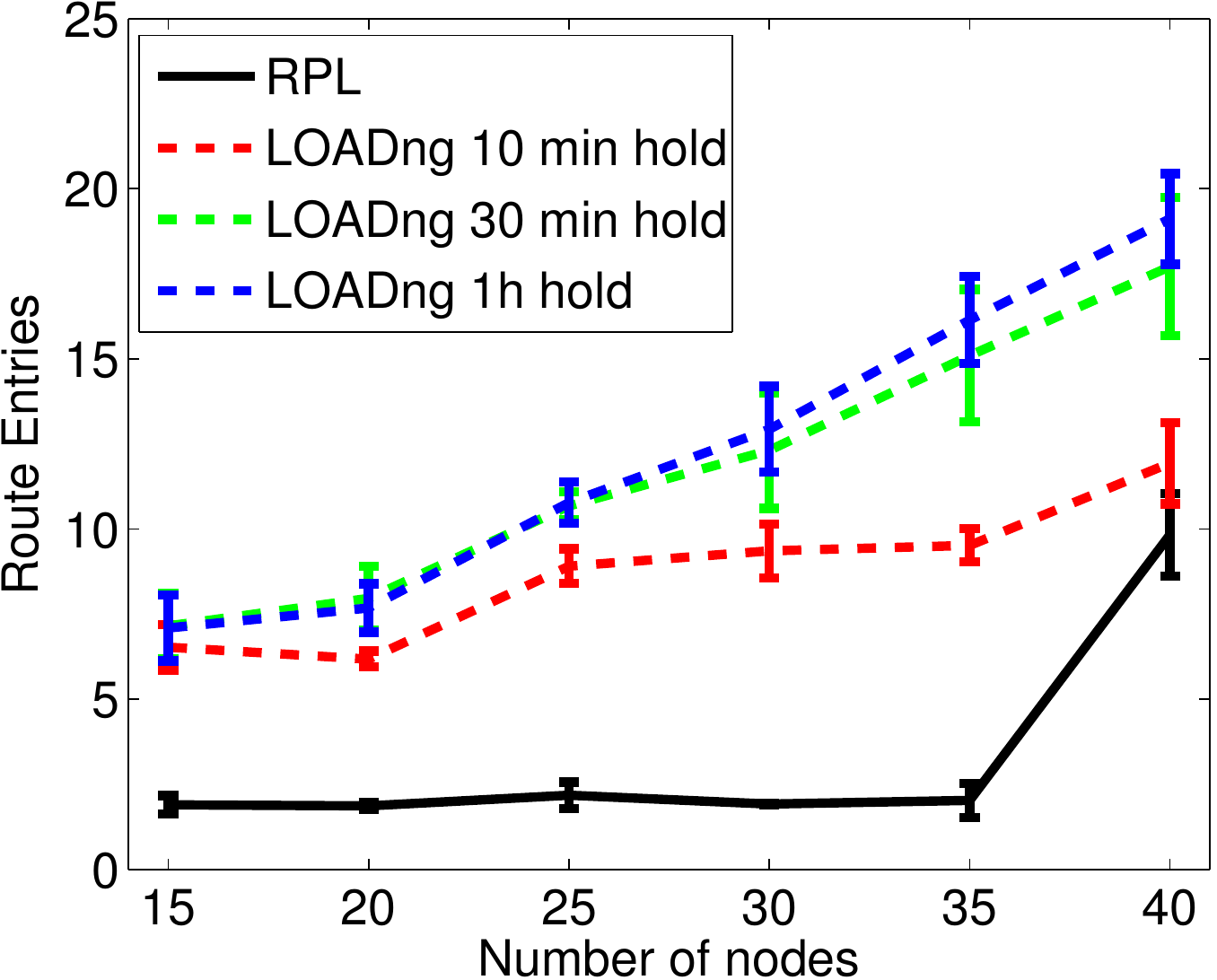}
\label{routingtable-increasing}
}
\subfigure[Control plane overhead (bytes).]{
\includegraphics[width=0.8\columnwidth, height=1.585in]{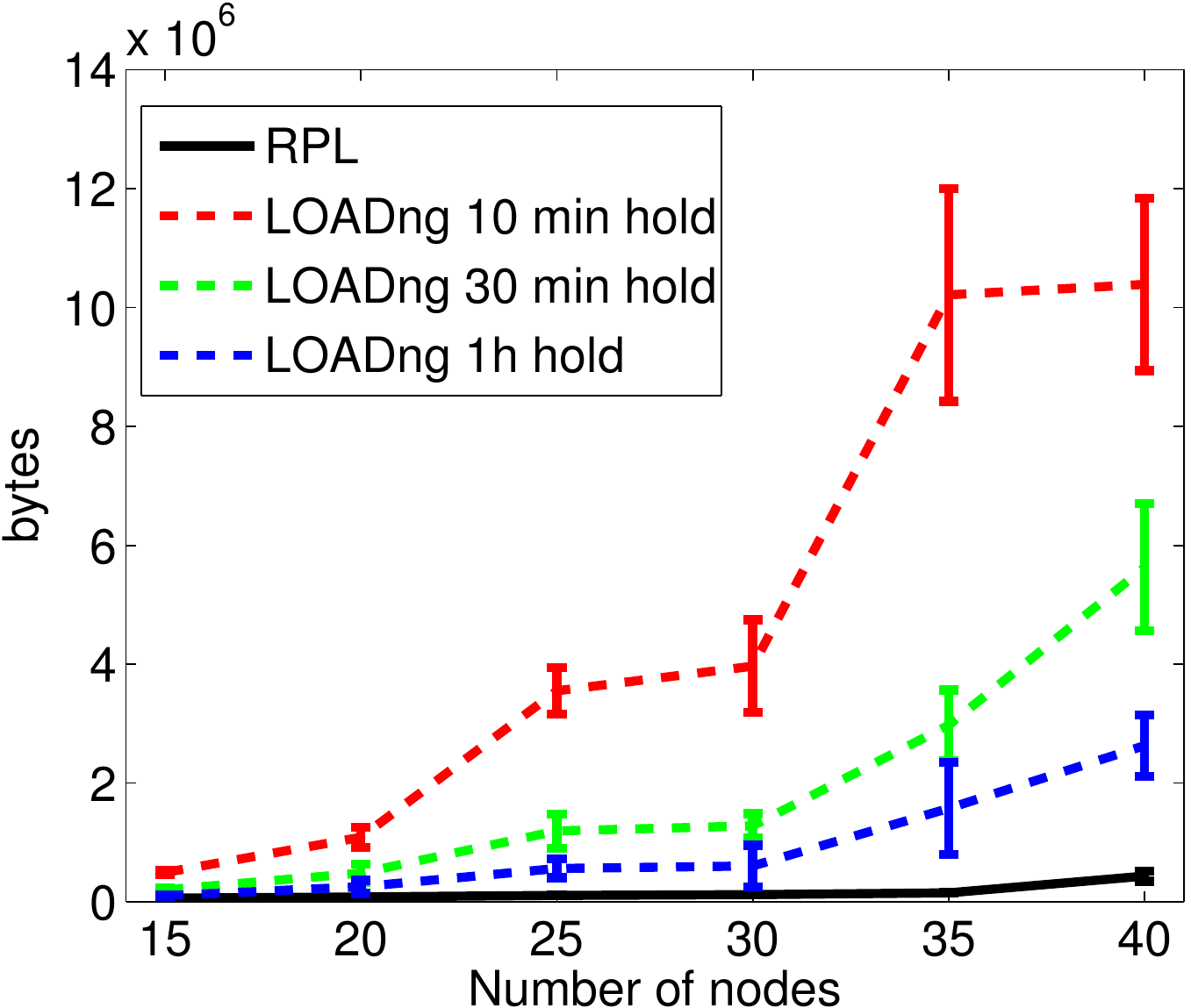}
\label{fig:overhead}
}
\caption{Memory and overhead for an increased number of nodes.}
\end{figure}

To see how the protocols scale for a larger network, we have studied
the average number of route entries as a function of an increasing number of
nodes (cf. Fig.
\ref{routingtable-increasing}). 
We can notice that the main consequence of having a long RHT
is a significantly increasing number of route entries. With
a shorter lifetime, the protocol scales better as the routes expire faster and
the average number of required entries slowly increases. For RPL, the average
number of routes slowly increases. 
However, it has a significantly higher number of entries for 40 nodes, a
consequence of the limited number of neighbor entries that nodes can
store. During the simulation time, we have observed oscillations in the RPL
graph as parent nodes were constantly overwritten by other neighbors so that a
number of nodes played the role of the preferred parent and stored downward
routes. This behavior may impact the RPL operation for more dense networks.  

\vspace{-0.2cm}\subsection{Overhead}
\label{overhead}

We have evaluated the routing overhead of the protocols as a function
of an increasing number of nodes (cf. Fig. \ref{fig:overhead}
presenting the overhead in bytes during the simulation time).
Although the previous work
\cite{rpl-load-clausen-paper} revealed a significantly higher 
overhead with RPL, our results show that RPL benefits from fairly low overhead
compared to LOADng. We evaluate the total overhead as a sum of the lengths of all control messages passed from the routing protocols to the layers bellow. Note that by doing so, we avoid taking into account the overhead introduced by ContikiMAC.

We can notice that the implementation of ContikiRPL results in a
lower overhead than observed previously \cite{rpl-load-clausen-paper}. 
As the RPL RFC under-specifies the DAO control message emission, the
overhead strongly depends on a given implementation. Furthermore, it is
important to note that our simulation scenario does not trigger any global
repairs of the RPL DODAG during the eight hours of simulation. 
In this way, our results give an idea about the performance of the two protocols
in the steady state. 
Thus, we observe that most of the RPL overhead appears at the beginning of
the simulation, reduced later by Trickle. 
The overhead of LOADng strongly depends on RHT: for a shorter RHT, nodes flood
more frequently. We can notice from the figure that the 
total amount of the LOADng overhead sharply increases with the number of nodes
due to the high density of nodes in the network, which  may however, be reduced
with an optimized flooding algorithm at the cost of a higher complexity.

The lifetime of battery-powered nodes  directly relates to the use of the
radio transceiver. More precisely, in case of the Tmote Sky platform, power
consumption due to the radio use is three orders of magnitude larger than that
due to  CPU processing \cite{tmote-sky}. Thus, the control plane
overhead and average number of hops of a protocol are the determining factors of
the expected node lifetime. Our study shows that the control overhead of LOADng
for short hold times does not scale well with dense deployments of smart home
applications. As 
we expect an increasing number of devices in home networks, this effect may
significantly impact the overall network lifetime.  
%
%
%

\section{Related Work}
\label{related-work}

Several authors analyzed the performance of RPL. An IETF draft
\cite{rpl-performance-draft} evaluates the protocol by considering several
routing metrics in real-life deployment scenarios. 
Nuvolone studied stability delays of RPL using OmNet++ \cite{nuvolone-thesis}. 
Clausen et al. studied the multipoint-to-point performance of RPL using NS2
\cite{clausen-m2p}. 
The authors \cite{clausen-optimized-broadcast} evaluated different optimized
broadcast techniques for the use in RPL
\cite{clausen-optimized-broadcast}. 
Ko et al. presented the implementation of RPL for TinyOS and discussed its
performance \cite{rpl-tiny}. 
Ben Saad et al. used the
Contiki framework to evaluate the performance of a PLC network serving as a LLN
backbone \cite{rpl-two-cases-studies}. 
Other authors studied the routing overhead and delay induced by ContikiRPL
implementation
\cite{rpl-performance}. 
Finally, an IETF draft \cite{rpl-experiences-draft} presented a thorough study of
several critical issues in the RPL protocol. 

Gomez et al. \cite{aodv-based-routing} compared various AODV-based routing
protocols, including LOAD and discussed design tradeoffs. 
Iliev et al. \cite{iliev-report} evaluated the route discovery delay as well as
the round trip time induced by the TinyOS implementation of LOAD in
real-life deployment scenarios. 

Herberg et al. presented a comparative performance study of RPL and LOADng in
case of bidirectional traffic using simulations in NS2
\cite{rpl-load-clausen-paper}. The paper shows a significantly larger control
overhead of RPL caused by the maintenance of downward routes. It also compares
the two protocols with an ideal routing protocol to show that RPL provides
near-optimal routes while LOADng results in a certain gap. 
As the RPL RFC does not specify the period or the mechanism to use for
maintaining downward routes, the study assumed an interval of 15 seconds. 
This choice is questionable, because it is the main cause of the high control
overhead of RPL. Furthermore, the application-layer scenario used for the
comparison is not the same for the two protocols. Thus, the question remains how
LOADng and RPL perform under the same scenario. 

\section{Conclusion}
\label{conclusion}

In the light of our study, RPL results in good overall performance for our
Home Automation scenario while LOADng could serve better sparse LLN deployments
with low-priority traffic in which the Route Hold Time can attain a  large
value. 
For Home Automation applications in which the response time is important, our
results suggest that LOADng is not the best candidate. 
However, it is important to recall that we have used a simple flooding scheme
and a better mechanism may reduce the control overhead. Furthermore, an
intelligent route storing algorithm may reduce memory requirements of
LOADng. These aspects are open research issues and their solutions will improve
the efficiency and scalability, but at the cost of a higher implementation
complexity. 

We have confirmed the previous findings that RPL provides shorter routes and a
smaller spanning tree depth compared to LOADng in our rather dense Home
Automation topologies. We have also showed that RPL delays stay small and
its overhead strongly depends on the implementation and a careful choice of
parameters. Moreover, RPL has shown better results and less memory requirements
than LOADng, but at the price of a higher implementation complexity. 

To sum up, several aspects remain interesting research challenges: reducing the
specification complexity of RPL, optimizing flooding and route storage of
LOADng as well as mixing the proactive and reactive approaches to maximize some
performance criteria for a given application domain. 

\section*{Acknowledgements}

This work was partially supported by the European Commission project CALIPSO
under contract 288879, the French National Research Agency (ANR) projects
ARESA2 under contract ANR-09-VERS-017, and IRIS ANR-11-INFR-016. 
Many thanks to Chi-Anh La for making available the LOADng implementation.

\bibliographystyle{IEEEtran} 
\bibliography{IEEEabrv,../Bibliography/general,../Bibliography/6lowpan_and_contiki,../Bibliography/rpl_load}

\end{document}